\documentclass[twocolumn,showpacs,showkeys,amsmath,amssymb,prb]{revtex4}
\usepackage{mathenv}
\usepackage{graphicx}
\usepackage{mathcomp}
\usepackage{ifthen}

\newcommand{\mathunit}[1]{\mathrm{#1}}
\newcommand{\size}[3][]%
{%
{#2}%
\ifthenelse{\equal{#2}{} \or \equal{#1}{}}{}{\cdot}%
\ifthenelse{\equal{#1}{}}{}{10^{#1}}%
\ifthenelse{\( \equal{#2}{} \and \equal{#1}{} \) \or \equal{#3}{}}{}{\, }%
\mathunit{#3}%
}%
\newcommand\diff{\mathrm{d}}
\newcommand\cbps{V_{\rm C}}
\newcommand\micro{\tcmu}
\newcommand\eV{\mathnormal{e}\:\!\!\mathunit{V}}
\newcommand{\main}{_{\rm m}}
\newcommand{\impty}{_{\rm t}}
\newcommand{\gate}{^{\rm g}}
\newcommand{\source}{^{\rm s}}
\newcommand{\drain}{^{\rm d}}

\newcommand\Vg{{V_{\rm g}}}
\newcommand\Vd{V_{\rm d}}

\newcommand\epV{\frac{\mathnormal{e}}{V}}
\newcommand\Cc{C_\mathrm{c}}
\newcommand\muB{\mu_\mathrm{B}}

\begin{document}

\title{Individual charge traps in silicon nanowires: 
Measurements  of location, spin and occupation number 
by Coulomb blockade spectroscopy}

\author{M.~Hofheinz}
\email{max.hofheinz@cea.fr}
\author{X.~Jehl}
\author{M.~Sanquer}
\affiliation{CEA-DRFMC, 17 rue yes Martyrs, F-38054 Grenoble cedex 9, France}
\author{G.~Molas}
\author{M.~Vinet}
\author{S.~Deleonibus}
\affiliation{CEA-LETI, 17 rue des Martyrs F-38054 Grenoble cedex 9, France}

\date{\today}

\begin{abstract}

We study anomalies in the Coulomb blockade spectrum of a quantum dot
formed in a silicon nanowire. These anomalies are attributed to
electrostatic interaction with charge traps in the device. A simple
model reproduces these anomalies accurately and we show how the
capacitance matrices of the traps can be obtained from the shape of
the anomalies. From these capacitance matrices we deduce that the
traps are located near or inside the wire. Based on the occurrence of
the anomalies in wires with different doping levels we infer that
most of the traps are arsenic dopant states. In some cases the
anomalies are accompanied by a random telegraph signal which allows
time resolved monitoring of the occupation of the trap. The spin of
the trap states is determined via the Zeeman shift.

\end{abstract}



\maketitle

\section{Introduction}
Single electron charges or spins are very appealing as logic bits,
either as ultimate classical bits or quantum bits if coherence is
used\cite{elzerman04,gorman05}. To read such bits, either quantum
point contacts or single electron transistors (SETs) are used. SETs
have indeed been used as very sensitive electrometers for the
(time-averaged) charge on a second quantum dot for over a decade
now.\cite{lafarge91,molenkamp95}. More recently the radio-frequency
SET technique\cite{schoelkopf98} was used to monitor the charge on the
second dot or to measure a current by electron
counting. \cite{lu03,fujisawa04,gustavsson06,bylander05}. This allows
to measure lower currents than standard measurements and gives access
to the full counting statistics of the current \cite{utsumi06}.

Such experiments are difficult because any device that involves
detection of few or single electron charges is subject to the dynamics
of surrounding charge traps. \cite{jung04} This is particularly
critical for metallic SETs. \cite{zimmerman97} For SETs based on the
very mature silicon CMOS technology the control of this offset charges
seems to be better. \cite{jehl02}

These charge traps are quantum dots whose presence or properties are
not controlled.  Typically they consist of defects on atomic
scale. Their sizes are therefore much smaller than what is possible
for lithographic quantum dots. If their positions, although being
random, can be limited to some zone, the charge traps are not
necessarily a nuisance but can useful. An example are flash memories
where the trend is to replace the lithographic floating gate by grown
silicon nanocrystals inside the gate oxide. They are grown in a layer
and have all the same distance from channel and gate electrode. Their
dynamics are therefore very similar. \cite{yano94,molas04b} Another
example are dopants in semiconductors. Efforts are made to control
their individual position in a silicon crystal. \cite{schofield03}
Indeed, besides the location, their properties are very uniform and
solid state quantum bits based on dopants in a silicon crystal
--individually addressed by gates and contacts-- were proposed as
solid state quantum bits \cite{loss98,kane98,hollenberg04}.  Silicon
is interesting as host material because the spin relaxation time can
be very long\cite{tyryshkin03} compared to GaAs.  The detection of
spins of individual traps in a silicon field effect transistor has
been recently reported using random telegraph noise
\cite{xiao03,xiao04}.  However, in this experiment the traps seem to
be in the oxide rather than in the silicon.

In this work, we use nanowire-based silicon transistors operated as
SETs at low temperature to detect the location, spin and occupation
number of individual charge traps, which we attribute to As dopant
states. They are capacitively coupled to the SET. Therefore they
induce anomalies in the otherwise very regular periodic oscillations
of the drain-source conductance $G$ versus gate voltage $\Vg$. We
compare the data with simulations obtained after solving the master
equation for the network formed by the main dot and the charge trap.

Not only the static time-averaged current is analyzed but also the
switching noise which appears near the degeneracy point in gate
voltage where the trap occupation number fluctuates.

Finally, a magnetic field was applied in order to probe the spin
polarization of the traps via their Zeeman shifts. As expected from
simple considerations \cite{shklovskii84}, we observed a majority of
singly occupied traps.

\section{Samples and Setup} 
\label{s:samples}
\label{s:measurement}

\begin{figure}
\includegraphics{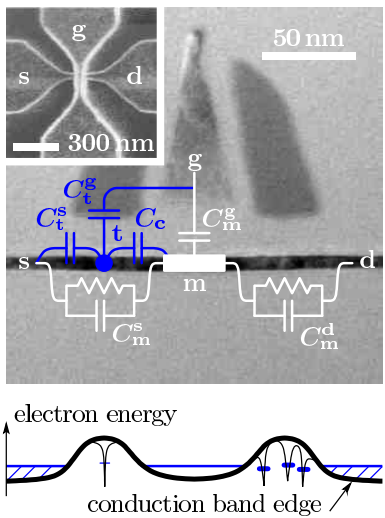}%
\caption{\label{scheme} (color online) Sample layout and electrical
model. The insert shows a top view of the sample before back-end
process obtained in a scanning electron microscope. The main image
shows a transmission electron micrograph (TEM) of a type B
sample along the silicon nanowire (black, the wire shown here is
thinner than in the samples used for measurements). Light gray regions
are silicon oxide. The darker region in the center is the polysilicon
gate with $\mathrm{Si_3N_4}$ spacers on both sides of it. Below, a
schematic energy diagram is drawn. The reduced doping level below the
spacers and the gate electrode creates a potential barrier, in the
middle of which a well is created by a positive gate
voltage. Conductance through the barriers separating the well from
source and drain occurs by tunneling through a chain of well connected
dopants (plotted in the right barrier). \cite{savchenko95} In more
isolated dopants (plotted in the left barrier) the number of charges
is well quantified. Such traps are the main concern of this
paper. Their interaction with the quantum well is mainly
electrostatic. We describe it with the lumped network superimposed to
the TEM.}%
\end{figure}

Samples are produced on $\size{200}{mm}$ silicon on insulator (SOI)
wafers with $\size{400}{nm}$ buried oxide and a boron substrate doping
of $\size[15]{}{cm^{-3}}$. The SOI film is locally thinned down to
approximately $\size{20}{nm}$ and a $\size{30}{nm}$ wide and
$\size{200}{nm}$ long nanowire is etched from it. A $\size{40}{nm}$
long polysilicon control gate is deposited in the middle of the wire
(see Fig.~\ref{scheme}). There are two layouts. Type A: The wires are
uniformly doped with As, above $\size[19]{}{cm^{-3}}$. The gate oxide
is $\size{4}{nm}$. Type B: The wires are first uniformly doped at a
lower level (As, $\size[18]{}{cm^{-3}}$), then, after deposition of
the gate electrode and $\size{50}{nm}$-wide $\mathrm{Si_3N_4}$ spacers
on both sides of it, a second implantation process increases the
doping to approximately $\size[19]{4}{cm^{-3}}$ in the uncovered
regions while the doping level stays low near the gate. In this layout
the gate oxide is 10 or $\size{24}{nm}$ thick, with a 2 or
$\size{4}{nm}$ thermal oxide and 8 or $\size{20}{nm}$ deposited
oxide. Most measurements are made on type B samples, and we use type A
samples mainly for comparison.

The measurements were performed in a dilution refrigerator with an
electronic base temperature of approximately $\size{150}{mK}$. We used
a standard 2-wire low frequency lock-in technique with low enough
voltage excitation to stay in the linear regime and a room temperature
current amplifier (gain $\size{100}{M\Omega}$). For time resolved
measurements a DC bias voltage was applied and current measured with a
$\size{10}{M\Omega}$ current amplifier (bandwidth $\size{10}{kHz}$)
followed by a $\size{33}{kHz}$ AD conversion.  For spin sensitive
measurements a superconducting magnet was used to apply an in-plane
magnetic field up to $\size{16}{T}$.

\section{Data}
\label{s:data}
\begin{figure}
\includegraphics{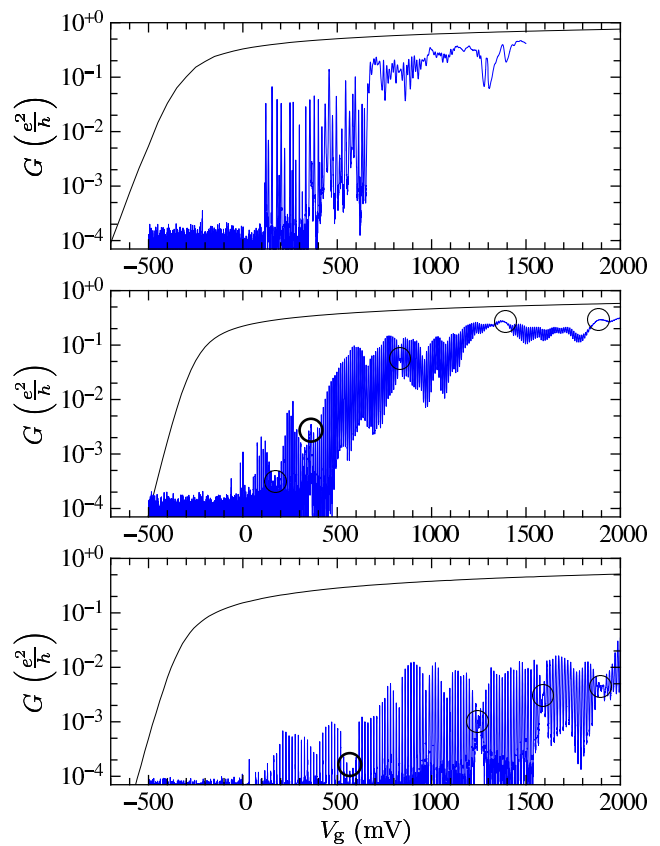}
\caption{\label{peakspacing} (color online) Drain-source conductance
versus gate voltage for 3 different samples. All samples have the same
width and gate length but the sample in the upper panel is of type A
with a $\size{4}{nm}$ gate oxide, while the samples in the lower
panels are of type B. The one of the middle panel has $\size{10}{nm}$
gate oxide, the one of the lower panel $\size{24}{nm}$.  The smooth
field-effect characteristics at room temperature (black lines) are
replaced by Coulomb blockade oscillations at base temperature (blue
curves). The period is determined by the surface area of the
nanowire/gate overlap. The Coulomb blockade oscillations in the upper
panel are irregular compared to the ones in the lower panel where only
some rare anomalies perturb the very regular spectrum. These anomalous
regions with reduced contrast and fluctuating peak spacing are
highlighted with circles. The anomalies marked with bold circles are
studied in detail in this work.}
\end{figure}

Figure \ref{peakspacing} shows typical $G(\Vg)$ plots.  At room
temperature our samples behave as classical (albeit not optimized)
$n$-channel MOSFETs. Below approximately $\size{20}{K}$ they turn into
single electron transistors with regularly spaced Coulomb blockade
resonances. The period $\cbps = \frac{e}{C\gate}$ of these
oscillations ($e$ is the absolute value of the electron charge) is
determined by the gate capacitance $C\gate$, which in turn can be
estimated from the gate/wire overlap and the gate oxide thickness.
For the sample with $\size{4}{nm}$ (A), $\size{10}{nm}$ (B),
$\size{24}{nm}$ (B) gate oxide the peak spacing is respectively
$\size{14}{mV} \pm \size{4}{mV}$, $\size{10.3}{mV}\pm \size{0.5}{mV}$,
$\size{15.3}{mV} \pm \size{0.8}{mV}$. This corresponds to a gate
capacitance of $\size{11}{aF}$, $\size{15.5}{aF}$,
$\size{10.5}{aF}$. For the type A samples the gate capacitance is in
good agreement with the simple planar capacitance estimation. For the
type B samples where the gate oxide thickness is of the same order as
the dimensions of the wire, the 3-dimensional geometry has to be taken
into account. The gate capacitance of the type B samples is increased
with respect to the type A samples because the flanks of the wire play
a more important role. A 3-dimensional numerical solution obtains a
good agreement with the measured capacitances.

The peak spacing statistics has already been measured and compared to
theory. \cite{boehm05} Here we focus on anomalous regions where the
conductance contrast is markedly reduced and a phase shift of the
Coulomb blockade oscillations occurs. This results in tails in the
Gaussian peak-spacing distribution. Such perturbations to the periodic
pattern are marked with circles in Fig.~\ref{peakspacing}. In the type
B samples with low doping, these perturbations occur only rarely (we
observe typically 3 to 5 per sample). In the unperturbed regions, the
height of the Coulomb blockade peaks shows long-range correlations. In
the type A samples with high doping level the perturbations are more
frequent and let the whole spectrum look irregular. (see top panel of
Fig.~\ref{peakspacing}) This suggests that the perturbations are
related to the doping.

\begin{figure}
\includegraphics{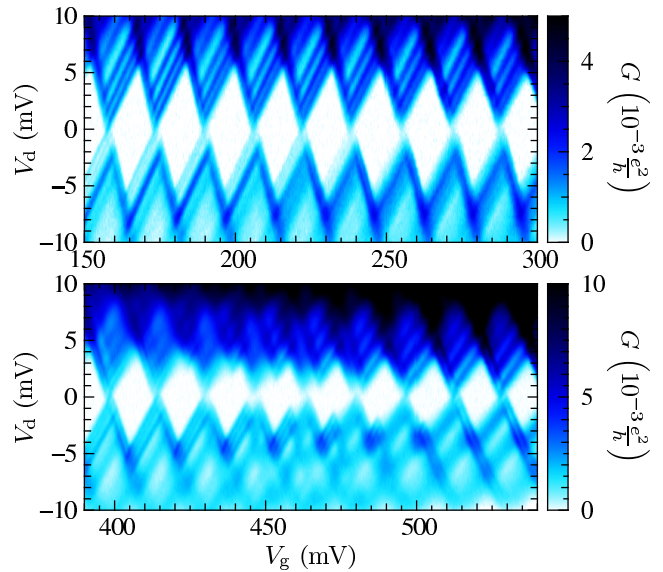}%
\caption{\label{rhomb-ex} 2D-plots of the measured drain-source
conductance versus gate and drain voltages in an unperturbed, very
periodic gate voltage range (upper panel), and in an anomalous region
where a charge trap is observed (lower panel). White areas correspond
to Coulomb blockaded regions (no detectable current). The lines inside
the conducting regions are not the excited states of the dot (the
spacing being too high and almost identical for all resonances). We
attribute these lines to additional conduction channels on the drain
side that open at higher bias (chains of well connected dopants lying
somewhat higher in energy than the drain Fermi level). Compared to the
measurements in Fig.~\ref{peakspacing}, the anomalous region has
shifted by $\size{50}{mV}$ in gate voltage after thermal cycling
between base and room temperature. We do not observe such shifts as
long as the sample is kept cold.}%
\end{figure}

In the measured stability diagram, i.e.\ the 2D plot of conductance
versus gate and bias voltages, the perturbations are even more visible
(see Fig.~\ref{rhomb-ex}).  In the perturbed regions additional teeth
appear in the Coulomb diamonds.

We develop a simple model based on a trap state located in the
vicinity of the quantum dot, and compare the simulation with the
experimental data.

\section{Model}
\label{s:model}
The quantum dot formed by the gate electrode in the middle of the wire
is separated from the source and drain reservoirs by a piece of
silicon wire containing only a few tens (type B) or hundreds (type A)
of dopants. In the type A samples these access regions extend from the
border of the gate electrode to the regions where the wire widens (see
Fig.~\ref{scheme}) and its resistance becomes negligible. In the type
B samples only the zones below the spacers contribute significantly to
the access resistance and the highly doped parts of the wire can be
considered as part of the reservoirs.

Electrons pass through these access regions by transport via the
dopant states.\footnote{Direct tunneling through a $\size{50}{nm}$
thick barrier leads to access resistances much higher than observed,
even for barrier heights of a few $\size{}{m\eV}$}. As the dopants are
distributed randomly and the coupling between them depends
exponentially on their distance, this coupling is distributed over a
wide range. Transport therefore takes place mainly through a
percolation path formed by well connected dopants\cite{shklovskii84}
while other dopant states are only weakly connected and their
occupation is a good quantum number (see Fig.~\ref{scheme}). We
attribute the anomalies in the Coulomb blockade spectrum to the
electrostatic interaction of the quantum dot with such a charge trap
formed by an isolated dopant site.

\begin{figure}
\includegraphics{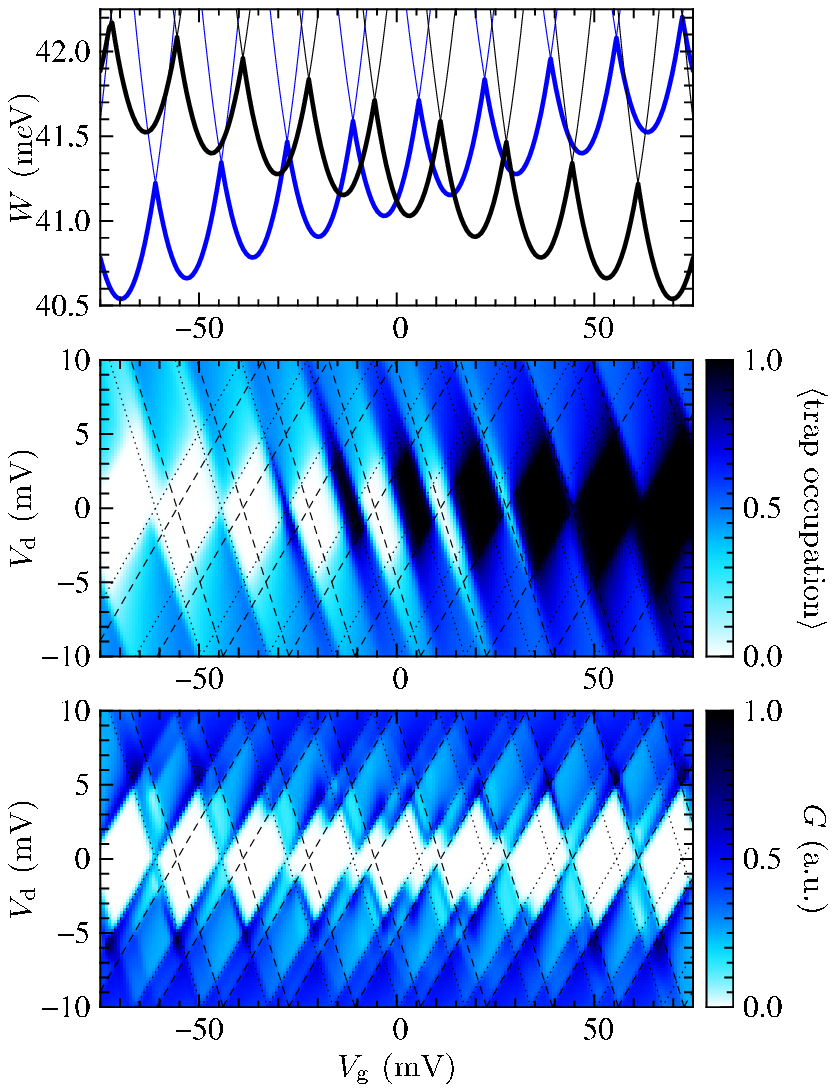}
\caption{\label{rhomb-sim}(color online) Numerical study of a trap
coupled to the source and to the main quantum dot, as sketched in
Fig.\ref{scheme}. Parameters: effective temperature: $T=\size{1}{K}$;
main dot: $C\main\gate = \size{60}{\epV}$, $C\main\drain =
C\main\source = \size{70}{\epV}$; trap: $C\impty\gate =
\size{0.045}{\epV}$, $C\impty\source = \size{2}{\epV}$,
$C\impty\drain=0$, $\Cc = \size{1}{\epV}$. The trap can either be
empty or charged with one electron. The upper panel shows the energy
for the different charge states at zero bias in function of gate
voltage. The blue parabolas are for empty trap, the black ones for
occupied trap. The thick blue and black lines indicate the ground
state of the main dot for respectively empty and occupied trap. The
middle panel shows the self-consistent mean occupation number of the
trap, the lower panel the resulting conductance through the dot. The
effect of the charge trap is to shift the Coulomb blockade diamonds of
the main dot depending on the charge in the trap. The dotted (dashed)
lines indicate the position of the diamonds when the trap is empty
(occupied). This result is in very good agreement with the
experimental data shown in the lower panel of Fig.~\ref{rhomb-ex}.}
\end{figure}

We model this with the lumped network shown in
Fig.~\ref{scheme}. Similar models have been considered in
Refs. \onlinecite{grupp01} and \onlinecite{berkovits05}. A small trap
(t) is capacitively coupled to source (s), gate (g) and to the main
dot (m). We note $C_i=C_i^{\rm s}+C_i^{\rm d}+C_i^{\rm g}$ and
$X_i=C_i^{\rm s}V_{\rm s}+C_i^{\rm d}V_{\rm d}+C_i^{\rm g}V_{\rm g}$
($i=\mathrm{m,t}$) After some calculation, the electrostatic energy of
the two dot system can be expressed as a function of the charges
$Q\main$ on the main dot and and $Q\impty$ in the trap.
\begin{equation}
W(Q\main,Q\impty)
=\underbrace{
\frac{\left(Q\main + \beta\impty Q\impty+ X\right)^2}{2 C}
}_{M(Q\main,Q\impty)} 
+ \underbrace{
\frac{\left(Q\impty + X\impty\right)^2}{2\left(C\impty+\Cc\right)}
}_{T(Q\impty)}
\label{elenergy}
\end{equation}
where $\Cc$ is the capacitive coupling between dot and trap,
$\beta\impty = \frac{\Cc}{C\impty+\Cc}$, $C = C\main + \beta\impty
C\impty$ and $X = X\main + \beta\impty X\impty$. For a small trap
($C\main < C\impty$) these renormalizations are weak: $C \approx
C\main$ and $X \approx X\main$. The problem is symmetric under
exchange of main dot and trap even though the expression in
Eq.~\ref{elenergy} is not. $W$ is plotted in the top panel of
Fig.~\ref{rhomb-sim}.

We focus on the structure of the Coulomb blockade conductance fixed by
Eq.~(\ref{elenergy}) and not on the exact value on the conductance
plateaus. Therefore we choose as simple as possible the following
parameters which are necessary for the simulation but do not affect
the structure of the conductance diagram.

We suppose all transmission coefficients to be constant, the ones
connecting the main dot to source and drain being 1000 times higher
than the ones connecting the trap to the main dot and source or
drain. Electrons can therefore be added or removed from the trap, but
their contribution to the total current through the device is
negligible. This contrasts with models of stochastic Coulomb blockade
\cite{ruzin92} or in-series quantum dots \cite{waugh95,rokhinson02}
where the current has to pass through both dots.

In terms of kinetic energy, we describe the main dot as metallic
(negligible single-particle level spacing $\Delta$, i.e.\ $\Delta \ll
kT$) and we consider only one non-degenerate energy level for the
trap. In source and drain we suppose a uniform density of states. We
assume fast relaxation of kinetic energy inside the dot and the
reservoirs, i.e.\ thermal distributions in the electrodes and the main
dot, even for nonzero bias voltage.  With these assumptions, the
transition rates of an electron in the main dot to the source or drain
reservoirs or from the reservoirs to the dot are proportional to the
auto-convolution of the Fermi function. The transition rates from or
towards the trap are directly proportional to the Fermi
function. \cite{grabert92}

The statistical probability for each state $(Q\main,Q\impty)$ of the
system can now be calculated by solving numerically the master
equation and gives access to the mean current through the system.

Results of such a numerical study are presented in
Fig.~\ref{rhomb-sim}. The middle panel shows the mean occupation of
the trap. On a large scale, the trap becomes occupied with increasing
gate voltage. In the central region of the figure however, whenever an
electron is added onto the main dot, the electron in the trap is
repelled and only later it is re-attracted by the gate electrode.
Inversely, the trap charge repels the charges on the main dot and the
Coulomb blockade structure of the main dot is shifted to higher gate
voltage when the trap is occupied (see lower panel). The two Coulomb
blockade structures for unoccupied and occupied trap are respectively
indicated by dotted and dashed lines in the middle and lower panel of
Fig.~\ref{rhomb-sim}.

This explanation is illustrated in terms of energy in the top panel of
Fig.~\ref{rhomb-sim}, which shows the energies for the different
charge configurations. The crossings of the blue (black) parabolas
give the positions of the Coulomb blockade peaks for empty (occupied)
trap. The shift between the crossings of the black parabolas with
respect to the crossings of the blue parabolas and the shift of the
dashed lines with respect to the dotted lines are due to the term
$\beta\impty Q\impty$ in $M(Q\main,Q\impty)$. Knowing that one Coulomb
blockade oscillation corresponds to a change of $e$ in $\beta\impty
Q\impty + X$, the shift due to $\Delta Q\impty= e$ is
\begin{equation}
\delta \Vg = \beta\impty \cbps 
\label{eq:shift}
\end{equation}
where $\cbps$ is the Coulomb blockade peak spacing of the main dot.

We will now determine the width of the anomaly in the Coulomb blockade
spectrum at low bias voltage. It is given by the gate voltage range
where the occupation of the trap oscillates at zero bias. In the top
panel of Fig.~\ref{rhomb-sim} this is the zone between the first and
the last crossing of the thick black line and the thick blue
line. First we calculate $\Delta M$, the difference of the ground
state energies for empty and occupied trap arising from the term $M$
in Eq.~\ref{elenergy}.  Then we calculate the change in gate voltage
necessary for $T(-e)-T(0)$ to exceed this difference.

$\Delta M$ reaches its extreme values when for one state of the trap
main dot is at a degeneracy point (the kinks in the thick lines),
where $M=\frac{(e/2)^2}{2C}$. For the other state of the trap the main
dot is a fraction $\beta\impty$ of a Coulomb blockade period away from
the degeneracy point and $M=\frac{e^2(1/2-\beta\impty)^2}{2C}$. The
extrema of $\Delta M$ are therefore $\pm
\frac{e^2}{2C}\beta\impty(1-\beta\impty)$.

The gate voltage dependence of term $T$ is given by $\alpha\impty =
\frac{1}{-e}\frac{\diff}{\diff \Vg} \left(T(-e)-T(0)\right) =
\frac{C\impty^{\rm g}}{C\impty+\Cc}$. Note that $\alpha\impty$ is the
long-range gate voltage lever arm of the trap over several Coulomb
blockade oscillations, where the charge of the main dot has to be
considered as relaxed with the source and drain Fermi
levels. $T(-e)-T(0)$ has to pass form
$+\frac{e^2}{2C}\beta\impty(1-\beta\impty)$ to
$-\frac{e^2}{2C}\beta\impty(1-\beta\impty)$ in order to toggle the
trap definitively. The width $\Delta \Vg$ of the anomaly is therefore
given by $-e \alpha\impty \Delta \Vg =
-2\frac{e^2}{2C}\beta\impty(1-\beta\impty)$ or
\begin{equation}
 n = \beta\impty(1-\beta\impty) \frac{\alpha\main}{\alpha\impty}
\label{eq:width}
\end{equation}
$n$ is the number of anomalous periods and
$\alpha\main=\frac{C\gate}{C}$ with $C\gate=C\main\gate+\beta\impty
C\impty\gate$ the gate voltage lever arm of the main dot.

\begin{figure}
\includegraphics{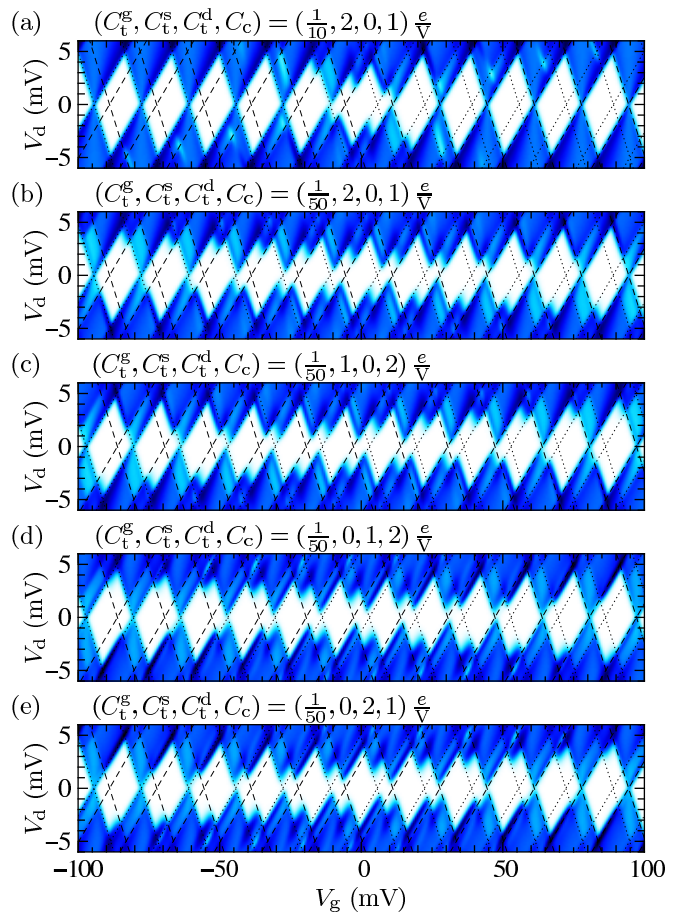}
\caption{\label{trap-pos} Calculated trap signatures for different
sets of parameters. (a) The trap is close to the source. (b) The
coupling to the gate electrode is reduced by a factor of 5. The
signature becomes wider. (c) The coupling to the source is reduced,
the coupling to the dot increased. (d) The trap is placed on the drain
side of the dot instead of the source side. (e) The coupling to the
dot is reduced, the coupling to the drain increased.}
\end{figure}
We have identified $\alpha\impty=\frac{C\impty\gate}{C\impty\! +
\!\Cc}$ and $\beta\impty=\frac{\Cc}{C\impty+\Cc}$ as parameters that
determine the structure of the trap signature. Both do not depend on
the absolute value of the trap's capacitances. Indeed, if one allows
only 0 or 1 electron in the trap, the absolute value of the trap
capacitances enters the problem only indirectly by modifying slightly
the capacitance matrix of the main dot and cannot be obtained in the
limit of a small trap. Our model contains therefore only 2 effective
parameters for the trap instead of 3 ($C\impty^{\rm g}$, $C\impty^{\rm
s}$, $\Cc$). All 3 parameters of the trap are only significant if the
trap can accommodate 2 or more electrons. In this case the spacing
between the anomalies gives access to the absolute values of the
trap's capacitances

Figure~\ref{trap-pos} illustrates the relation between the trap's
capacitance matrix and its signature.  If teeth of constant width for
all anomalous resonances are visible at the positive slope of the
Coulomb blockade diamonds, the trap is on the source side of the
dot. If they are visible at the negative slope, the trap is on the
drain side. The width of the teeth depends on $\beta\impty$, the width
of the anomalous region essentially on $\alpha\impty$ (for
$\beta\impty$ close to $\frac{1}{2}$ where the anomalies are well
visible).

\section{Position and nature of the traps}
\label{s:discussion}
As an illustration, from the lower panel of Fig.~\ref{rhomb-ex} we
infer $\alpha\impty \approx 0.015$ and $\beta\impty \approx
0.3$. These are the actual parameters that have been chosen for the
simulation in Fig.~\ref{rhomb-sim} and the lower panels of
Fig.~\ref{rhomb-ex} and Fig.~\ref{rhomb-sim} are indeed very
similar. As for all impurities we observed, $\alpha\impty$ is
small. This is what we expect for a trap inside the silicon wire. The
coupling to the gate electrode is much weaker than the coupling to the
main dot or the source electrode because the dielectric constant of
the oxide barrier ($\epsilon_{\rm SiO_2} = 4$) is much smaller than
that of bare silicon ($\epsilon_{\rm Si} = 12$), which in addition is
enhanced near the insulator--metal transition. \cite{imry82}

Positions of the trap outside the wire can be ruled out. Traps located
deep inside the oxide can be excluded because their transmissions
would be too weak to observe statistical mixing of occupied and
unoccupied trap states during our acquisition time below
$\size{1}{s}$. Similar devices including intentional silicon
nanocrystals at the interface between thermal oxide and deposited
oxide have been studied in views of memory
applications. \cite{molas04b,molas05} The measured lifetime of charges
in the nanocrystals exceeds $\size{1}{s}$ by orders of magnitude
already at room temperature and at low temperature gate voltages of
about $\size{5}{V}$ have to be applied in order to toggle the charge
in the nanocrystals. The traps must therefore be inside the Si wire or
at its interface with the oxide. But the interface traps are unlikely. We
estimate their density to be smaller than $\size[11]{}{cm^{-2}}$,
corresponding to a few units per sample. As they are distributed
throughout the entire band gap it is very unlikely to observe several
of them in the small energy window $\alpha\impty (\Vg^\mathrm{max} -
\Vg^\mathrm{min}) \approx \size{30}{m\eV}$ that we scan in our
measurement. The most likely traps are therefore defects in the
silicon wire or As donor states. Given the volume of the access
regions under the spacers and the doping level $N_\mathrm{D}$, there
are around 70 donor states under the spacers in devices of type B. We
estimate the width of the impurity band to be
$\frac{e^2}{\epsilon_0\epsilon_r N_\mathrm{D}^{-1/3}} \approx
\size{150}{m\eV}$. \cite{shklovskii84} One should therefore expect
around 15 dopants in the energy window. Typically we record 3 to 5
anomalies. Indeed we do not expect to observe anomalies for all
dopants because the charge on well connected dopant sites is not
quantified and, according to our model, dopants very close to the dot
($\beta\impty \approx 1$) or to the reservoir ($\beta\impty \approx
0$) produce very small anomalies.

In the type A samples the doping level in the access regions is more
than 10 times higher than in the type B samples. The whole Coulomb
blockade spectrum should therefore be anomalous. Indeed, the spectrum
is much less regular (see Fig.~\ref{peakspacing}) than for the type B
samples, especially for low gate voltage, but we cannot distinguish
signatures as clear as in the type B samples. This is consistent
because in the type A samples the mean distance between impurities is
less than $\size{3}{nm}$ and they are too well connected for the
charge on them to be well quantified. In other words, the wire is very
close to the insulator--metal transition. Our doping level is in fact
already higher than the bulk critical As concentration $N_{\rm c} =
\size[18]{8.6}{cm^{-3}}$. \cite{castner75,shafarman89}

We have deduced that the observed traps lie inside the wire. The
position of the trap along the wire can also be determined. Traps on
the source side and the drain side of the dot can be distinguished
(see Fig.~\ref{trap-pos}, the teeth of constant width $\delta \Vg$
appear on the positive slope of the diamonds in case of a trap on the
source side of the dot and on the negative slope in case of a trap on
the drain side) and the parameter $\beta\impty$ gives the ratio
between the capacitances towards the main dot and the source (or
drain) electrode. As the dielectric constant of the wire is much
higher than the surrounding silicon oxide, this ratio can be
translated linearly to a position in direction of the wire. In the
example of Fig.~\ref{rhomb-ex} with $\beta\impty \approx 0.3$ we would
expect the impurity to be located $\frac{2}{3}$ on the way from the
dot (border of the gate electrode) to the source reservoir (source
side border of the spacer).

\section{Time-resolved occupation number}
\label{s:switching}

\begin{figure}
\includegraphics{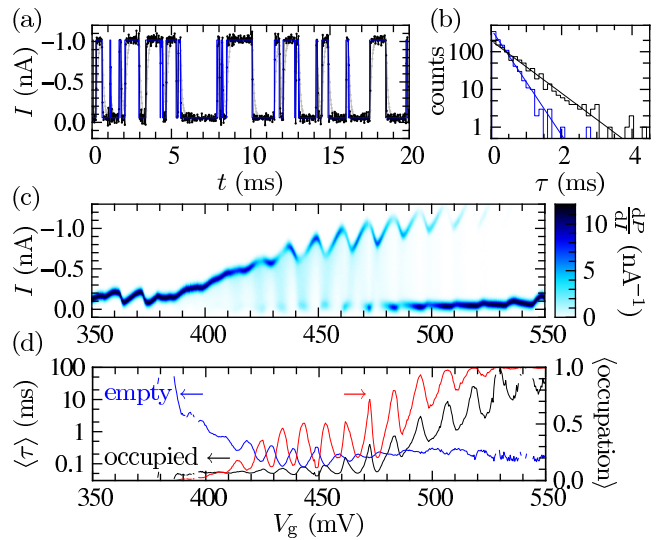}
\caption{\label{switching1}(color online) Analysis of a trap signature
with switching. Sample of type B with $\size{24}{nm}$ gate oxide. The
width of the wire is $\size{80}{nm}$ instead of $\size{30}{nm}$.
(a)~A RTS trace taken at $\Vg = \size{500}{mV}$, $\Vd =
\size{-6}{mV}$. Light gray trace: raw data. Black trace: data after
compensation of the time constant of the current amplifier. Blue line:
fitted signal.  (b)~Histograms of the times passed in the weak current
state (occupied trap, black) and the $\size{-1}{nA}$ state (empty
trap, blue). The time constants (averages of these times) are
$\size{0.31}{ms}$ and $\size{0.62}{ms}$. The corresponding exponential
distributions (straight lines) fit well the histograms. (c)~Current
histogram at $\Vd = \size{-6}{mV}$. The nonzero density between the
two current levels is due to the finite rise time. The current for
unoccupied trap is always higher than for occupied trap. (d)~Time
constants of the empty and occupied levels in function of gate voltage
and occupation number of the trap.}
\end{figure}

In the preceding sections we assumed charge traps with changing mean
occupation number to explain anomalies in the mean conductance through
a Coulomb blockaded quantum dot. Yet the measurements of the mean
current have not allowed us to measure the occupation number of the
trap directly. But the currents through the main dot differ for empty
and occupied trap because the position of the Coulomb blockade
resonances is shifted, and at the anomalies where the occupation
number of the trap is different from 0 and 1, the fluctuations of the
occupation number should create a random telegraph signal
(RTS)\cite{kirton89,xiao03,xiao04,buehler04} in the current through
the main dot.

Indeed, we frequently observe strongly increased current noise near
the anomalies, especially at low gate voltage. However, at most
anomalies at higher gate voltage we do not observe a clear increase
of the noise level. This indicates that changes in the trap state
occur at frequencies far beyond $\size{10}{kHz}$, the bandwidth of our
measurement. Indeed, for charge traps formed by dopants we would
expect the transmission rates of the trap to be of the same order
as for the main dot, where the transmission occurs also through dopant
states. The excess noise of the trap should therefore be comparable to
the shot noise of the quantum dot, which does not emerge from the noise
floor of the current amplifier. But much smaller transmissions should
also be possible as the dopants are distributed randomly and the
transmission rate depends exponentially on the distance between them.
In some cases we observe clear RTS with time constants larger than
$\size{100}{\micro s}$. An
example is given in Fig.~\ref{switching1}(a). The distribution of the
times spent in the two states follows the exponential distribution
expected for a RTS (see Fig.~\ref{switching1}(b)).

The color plot of the current distribution in Fig.~\ref{switching1}(c)
shows the evolution of the the two current levels (dark lines with
high probability) with gate voltage. Above $\size{380}{mV}$ the two
levels are very different. This difference is most likely due to
electrostatic interaction of the trap and the current path through the
barrier: depending on the state of the trap, the dopants through which
the main part of the current flows are well or poorly aligned in
energy. The fact that the current levels never cross simplifies
greatly the assignment of the high and low current levels to the
states of the trap. The high current trace being most likely at low
gate voltage and the low current trace being most likely at high gate
voltage allows to attribute the high current to empty trap and the
low current to occupied trap.

\begin{figure}
\includegraphics{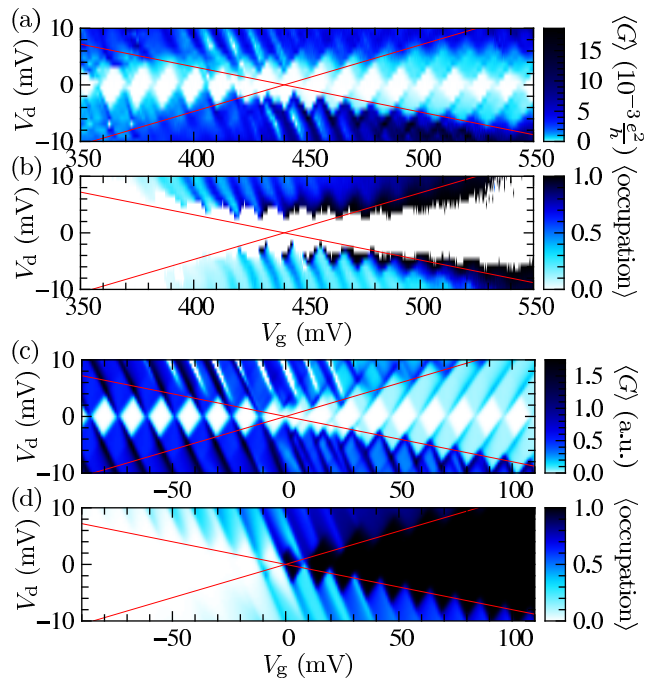}
\caption{\label{switching2}(color online) Comparison of measured
occupation number and simulation. Same trap as in
Fig.~\ref{switching1}. (a)~Mean differential conductance obtained by
numerical derivation of the mean current. (b)~Occupation of the trap
obtained from the duty cycle of the RTS signal. Regions where no clear
RTS could be detected are left white. (c) and (d) Simulation with the
following parameters: main dot: $C\main\gate = \size{80}{\epV}$,
$C\main\source = \size{60}{\epV}$, $C\main\drain=\size{100}{\epV}$;
trap: $C\impty\gate = \size{0.08}{\epV}$, $C\impty\source =
\size{0.6}{\epV}$, $C\impty\drain=0$, $\Cc = \size{1}{\epV}$. In units
of the drain--dot barrier transmission, the source--dot barrier
transmission is 10 for empty trap and $\frac{1}{10}$ for occupied
trap, the source--trap barrier transmission $\frac{1}{1000}$ and the
trap--dot barrier transmission $\frac{1}{3000}$.  }
\end{figure}

The time constants of the empty and occupied state are plotted in
Fig.~\ref{switching1}(d). Consistently with panel (c), the time
constant for the empty trap decreases with gate voltage while the time
constant for the occupied trap increases. Superimposed with this slow
change there are oscillations with a period of $\size{12}{mV}$, the
peak spacing of the main dot. This oscillation is even more prominent
in the mean occupation number given by $\frac{\tau_\mathrm{occupied}}{\tau_\mathrm{occupied}+\tau_\mathrm{empty}}$.
As explained in section~\ref{s:model} for the case of low bias, this
oscillation is due to the discrete charge on the main dot
which cycles the trap several times between empty and occupied state. It is
not observed in RTS in larger devices without Coulomb
blockade\cite{kirton89}. At high bias ($\Vd > \frac{e}{C}$) only an
oscillation of the occupation probability remains of this
cycling. This can be seen in Fig.~\ref{switching2}(b) and (d) where
the occupation probability for different bias voltages is compared
with simulation. As in Fig.~\ref{rhomb-sim}, the oscillations in
Fig.~\ref{switching2}(b) and (d) are aligned parallel to the negative
slopes of the Coulomb blockade diamonds indicating that the trap is on
the source side of the dot.

RTS (i.e.\ current through the trap) only occurs when the trap
is in the bias window.  For large gate and bias voltage excursions
where the charging energy of the main dot is negligible, the main dot
can be considered as part of the drain reservoir. The zone where the
trap is in the bias window is then delimited by slopes
$\frac{C\impty\gate}{C\impty}$ and $-\frac{C\impty\gate}{\Cc}$
(indicated by straight lines in Fig.~\ref{switching2}), just as for a
single quantum dot. These slopes give a more straightforward access to
the parameters $\alpha\impty$ and $\beta\impty$.

The mean occupation of the trap is higher for positive drain voltage
than for negative drain voltage indicating a higher transmission rate
of the trap towards source than towards the main dot.

In Fig.~\ref{switching2}(c) and (d) we try to reproduce panels (a) and
(b). For this simulation we reduce by a factor of 100 the transmission
of the source barrier of the main dot when the trap is occupied. This
reproduces the lines of reduced differential conductance at positive
drain voltage (compare Fig.~\ref{switching2}(a) and (c)). In the
simulation the oscillations of the trap occupation decay more rapidly
with bias voltage than in the measurement. This could be related to our
approximation of a thermal distribution of kinetic energies in the
main dot, which is certainly not accurate at high bias voltage.

Charge traps are generally believed to be not only responsible
for RTS noise but also for $1/f$ noise in SETs\cite{jung04} and
decoherence\cite{itakura03}. These interpretations imply a large
number of traps with small influence on the device (in our model
$\beta\impty \approx 0$). Such traps could be dopants in the
reservoirs or the substrate.

\section{Spin} 

\label{s:spin}

The spin of the trap state leads via the Zeeman energy under magnetic
field to a gate-voltage shift of the trap signature of:
\begin{equation}
e \alpha\impty \frac{\partial \Vg}{\partial B} 
= g \mu_\mathrm{B} \Delta S_\mathrm{z}
\label{zeeman}
\end{equation}
$\muB$ is the Bohr magneton, $\Delta S_\mathrm{z}$ the change in spin
quantum number of the trap state in direction of the magnetic field
when an electron is added to the trap. It can take the values $\pm
\frac{1}{2}$. If there are already electrons in the trap higher
changes are also possible, but they imply spin flips and such
processes are therefore expected to be very slow. \cite{weinmann95}
The Land\'{e} factor $g$ for impurities in Si and SiO$_2$ has been
measured by electron spin resonance. \cite{lenahan98} The observed
renormalizations are beyond the precision of our measurements,
therefore we take $g=2$. The gate-voltage lever-arm of the trap states
$\alpha\impty$ is very weak as we have shown above. The Zeeman shifts
should therefore be strong.

\begin{figure}
\includegraphics{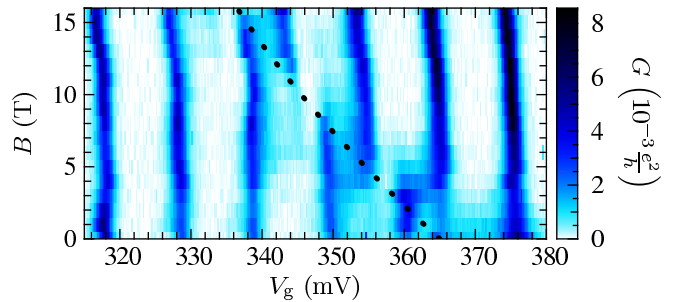}
\caption{\label{spin1}(color online) Shift of a trap signature with
magnetic field. The dotted line indicates the Zeeman shift expected
for a trap state being occupied by a first electron. It depends on the
gate-voltage lever-arm which in turn is determined by the width of the
signature. This prediction of the Zeeman shift follows exactly the
observed shift.}
\end{figure}

Indeed, the magnetic field clearly shifts the trap signature in
Fig.~\ref{spin1} to lower gate voltage. In order to identify the shift
as the Zeeman effect, we compare it quantitatively with the prediction
of our model. The shift of the resonances due to the trap is half the
peak spacing, so $\beta\impty = \frac{1}{2}$ (see
Eq.~\ref{eq:shift}). The lever arm for the main dot is for this gate
voltage $\alpha\main = 0.26$ and the width of the trap signature
varies from 2.5 periods without magnetic field to 1.5 periods at
$\size{16}{T}$. This implies a gate-voltage lever arm for the trap of
$\alpha\impty = 0.026 \cdots 0.043$ (see Eq.~(\ref{eq:width})) which we
interpolate as a linear function of magnetic field. The dotted line
in Fig.~\ref{spin1} is obtained if we put this lever-arm and
$S_\mathrm{z}=-\frac{1}{2}$ in Eq.~(\ref{zeeman}). It is in very good
agreement with the measured shift and confirms our model.  The
increase of the lever arm with magnetic field could be explained as
follows. In the access regions the nanowire is close to the
metal-insulator transition and the dopant states strongly increase the
dielectric constant\cite{castner75}. Under magnetic field they
shrink\cite{shklovskii84}, reducing the localization length and the
dielectric constant in the wire. Therefore the coupling towards the
main dot and the reservoir decreases while the gate capacitance
dominated by the oxide capacitance remains unaffected.

We observe such Zeeman shifts in the majority of our samples. In most
cases the trap signature shifts to lower gate voltage as in
Fig.~\ref{spin1}. This is what we expect for isolated traps occupied
with one electron. When a trap state is occupied with a second
electron it has to occupy the energetically less favorable state whose
energy is increased by the Zeeman effect. This leads to a shift
towards higher gate voltage under magnetic field. Although isolated
As-donor sites in Si can only be occupied by one \footnote{Arsenic
donors can be populated with 2 electrons but the second electron is so
weakly bound that in the scale of our devices it can be considered as
delocalized. See for example Ref.~\onlinecite{shklovskii84}} electron
due to Coulomb repulsion, clusters of two donors could contain two or
more electrons. \cite{bhatt81} For not too high doping levels clusters
should however be rare. Accordingly we observe much less shifts to
higher than to lower gate voltage. In devices based on similar
technology Xiao \emph{et al.} observed that all shifts occurred to
higher gate voltage\cite{xiao03,xiao04} indicating doubly occupied
traps. With precise measurements of the Land\'{e} factor they located
the traps inside the oxide. This difference also supports that the
traps in our device are not located in the oxide but inside the
silicon wire.

\section{Perspectives} 
Dopant states in silicon could provide very scalable solid state
quantum bits, based on charge, electron spin or nuclear spin. But it
is still very difficult to control their position individually. On the
other hand, with several gate electrodes one could imagine to select
suitable dopants out of large number of randomly distributed
dopants. In this context we have presented how the capacitance matrix
of charge traps near a small silicon single electron transistor can be
determined and we showed how the gate-voltage dependence of the
occupation is related to the spin of the trap state and that the
charge in these traps can be read out. These charge traps are
attributed to arsenic dopant states. At a doping level of
$\size[18]{}{cm^{-3}}$ we observe several well isolated dopant states
per device as well as percolation paths of well connected dopants
linking the main quantum dot to the reservoirs. In similar geometries
with multiple gate electrodes the coupling between the dopants could
be tuned by changing their alignment in energy with the well connected
dopants. Such randomly distributed dopants are probably more suited
for electron spin quantum bits than for charge quantum bits where two
dopant sites with small distance are necessary. In this perspective we
are working on measurement of the coherence time of the electron spin in the
observed traps. Together with the excellent stability in time as well
as its full compatibility with CMOS technology our system could be a
good basis for scalable quantum bits.

\section*{ACKNOWLEDGMENTS}
This work was partly supported by the European Commission under the
frame of the Network of Excellence ``SINANO'' (Silicon-based
Nanodevices, IST-506844).



\end{document}